\documentclass[a4paper,11pt]{article}
\usepackage{latexsym}
\usepackage{amsmath}
\usepackage{amssymb}
\usepackage{graphicx}
\usepackage{wrapfig}
\begin{document}


\noindent

\noindent
SAGA-HE-207-04

\bigskip

\centerline{\bf Effective Hadron Masses and Effective Interactions in Nuclear Matter}

\bigskip

\centerline{H. Kouno$^{1}$, Y. Horinouchi$^{1}$ and K. Tuchitani$^{2}$}

\noindent
$^1${ Department of Physics, Saga University, Saga 840-8502, Japan}

\noindent
$^2${ Saga High School of Technology, Saga 840, Japan}

\bigskip

\centerline{\bf Abstract}

\noindent
Relation among effective hadron masses, effective interactions and equation of state are studied using a generalized mean-field theory which includes the implicit and the explicit density dependence of the effective masses and couplings. 
We found that we can make the effective $\omega$-meson mass smaller and the equation of state softer at the same time if the $\omega$-meson mean field is proportional to the baryon density. 
In this case, there is a simple and exact relation between the effective $\omega$-meson mass and the effective $\omega$-nucleon coupling. 
Because of this relation, the effective $\omega$-nucleon coupling automatically decreases if the effective $\omega$-meson mass decreases. 
Consequently, the equation of state becomes softer. 
A trial to embed the QCD sum-rule results into the hadron field theory is also shown. 

\bigskip



\section{Introduction}

Changes of the hadron masses and couplings are much interested in the hadron and nuclear physics. \cite{rf:Brown,rf:Hatsuda,rf:Hatsuda2}
Especially, the vector meson mass reduction attracts a great deal of attention because it is related to the chiral symmetry restoration. 
Because of its short life time, the reduction of the $\rho$-meson mass is expected to be a signal of the hot and dense matter which may be produced in the high-energy heavy ion collisions. \cite{rf:Tserruya}  

On the other hand, the effective $\omega$-meson mass is important for the nuclear structure. 
The results of the relativistic Brueckner Hartree Fock calculation \cite{rf:Brockmann,rf:deJong} is well described by the $\omega$-meson self-interaction \cite{rf:Bodmer,rf:Sugahara} or the decrease of the $\omega$-nucleon couplings in medium. \cite{rf:Brockmann2} 
In the paper \cite{rf:Tuchitani}, Tuchitani et al. showed analytically that the enhancement of the effective $\omega$-meson mass caused by the $\omega$-meson self-interaction softens the equation of state (EOS). 
They also showed that the reduction of the effective $\omega$-nucleon coupling is related with the enhancement of the effective $\omega$-meson mass if it is caused by the effective multi $\omega$-nucleon interaction. 
However, it seems that such an enhancement of the effective $\omega$-meson mass is not consistent with the chiral symmetry restoration in medium. 
It is reported that the reduction of the effective vector meson mass makes the EOS stiffer. \cite{rf:Weber,rf:Hyun} 

In the paper \cite{rf:Tuchitani}, they have used the generalized mean field theory which includes the higher-order corrections, which is regarded to be induced by vacuum fluctuation, in the higher-order effective interactions. 
However, for the density fluctuation, their model include only the simple Hartree contributions. 
In this paper, we extend their model and include the higher-order effects in the density fluctuations as the explicit density dependence of the effective hadron masses and the effective couplings. \cite{rf:Fuchs} 
Using the more generalized mean-field theory, we study the relation among the effective hadron masses, the effective couplings and the EOS of symmetric nuclear matter. 
We found that, in many case, the reduction of the effective $\omega$-meson mass is followed by the enhancement of the effective $\omega$-nucleon coupling and the EOS becomes stiffer. 
However, we can make the effective $\omega$-meson mass and the effective $\omega$-nucleon coupling smaller at the same time, if we require that the value of the $\omega$-meson mean field is proportional to the baryon density. 
In that case, there is a simple relation between the effective $\omega$-meson mass and the effective $\omega$-nucleon coupling. 
Because of this relation, the effective $\omega$-nucleon coupling automatically decreases if the effective $\omega$-meson mass decreases. 
Consequently, the EOS becomes softer. 


We remark that the density dependence of the effective masses and couplings may be explained by the underlying quark physics rather than the hadron field theory itself. 
We may be able to embed the quark physics results in the hadron field theory through the density dependence of the effective masses and couplings. 
In this paper, we try to embed the results of QCD sum rule \cite{rf:Hatsuda,rf:Cohen} in the hadron field theory. 
In the viewpoint of the quark physics, it may be natural that the value of the $\omega$-meson mean field is proportional to the baryon density. 
\cite{rf:Kunihiro} 

This paper is organized as follow. 
In Sec. 2, we generalize the relativistic mean field theory to include the explicit density dependence \cite{rf:Fuchs} as well as the higher order interactions. \cite{rf:Tuchitani} 
We show that the density evolution equation for the effective potential plays an important role to determine the effective masses and couplings in the theory. \cite{rf:Berges} 
In Sec. 3, an analytical expression for the incompressibility of nuclear matter is shown. 
It is also shown that there is a simple relation between the effective $\omega$-meson mass and the effective $\omega$-nucleon coupling if the value of the $\omega$-meson mean field is proportional to the baryon density. 
In Sec. 4, we numerically investigate the relation among the effective $\omega$-meson mass, the effective $\omega$-nucleon coupling and the EOS. 
We found that we can make the effective $\omega$-meson mass smaller and the EOS softer at the same time if the value of the $\omega$-meson mean field is proportional to the baryon density. 
We also try to embed the $\omega$-meson mass reduction, which is predicted by the QCD sum-rule, \cite{rf:Hatsuda} into hadron field theory. 
In Sec. 5, we discuss the relation between the nucleon mass reduction predicted by the QCD sum-rule \cite{rf:Cohen} and the nucleon-nucleon interactions. 
We found that thermodynamical consistency is very important for studying 
the relation between the effective hadron masses and the effective interactions because the consistency condition is nothing but the evolution equation for the effective potential at finite density. \cite{rf:Berges} 


\section{Effective Lagrangian}

We start with the following effective Lagrangian of $\sigma$-$\omega$ model \cite{rf:Walecka,rf:Serot} which includes the higher-order interactions \cite{rf:Tuchitani} and the explicit density dependence. \cite{rf:Fuchs} 
\begin{eqnarray}
L&=&\bar{\psi}\left[\gamma^\mu\{i\partial_\mu+\Sigma_\mu (\sigma, \omega, S,V) -\{m+\Sigma_{\rm s}(\sigma ,\omega, S, V)\}\right]\psi
\nonumber\\
&+&{1\over{2}}\partial^\mu\sigma\partial_\mu\sigma
-{1\over{4}}F_{\mu\nu}F^{\mu\nu}
\nonumber\\
&-&U_{\rm M}(\sigma,\omega,S, V);
\nonumber\\
&F_{\mu\nu}&=\partial_\mu\omega_\nu -\partial_\nu\omega_\mu,~~~~~
\label{eq:E1}
\end{eqnarray}
where $\psi$, $\sigma$, $\omega_\mu$ and $m$ are the nucleon field, the $\sigma$-meson field, the $\omega$-meson field and the nucleon mass, respectively. 
The $S$ and $V$ are auxiliary variables which are equivalent to the scalar density $\rho_{\rm s}$  and the baryon density $\rho_{\rm B}$ of nucleons, respectively. 
The $\Sigma_{\rm s}$, $\Sigma_\mu$, and $U_{\rm M}$ are functions of $\sigma$, $\omega^\mu$, $S$ and $V$.   
The $\Sigma_{\rm s}$ and $\Sigma_\mu$ are the meson-nucleon interactions, while $U_{\rm M}$ is the mesonic potential which includes the meson mass terms, the $\sigma$-meson self-interaction, the $\omega$-meson self-interaction and so forth.
We emphasize that $U_{\rm M}$ also include the explicit density dependence through $S$ and $V$.  

Starting from the effective Lagrangian (\ref{eq:E1}), we calculate the density effects in nuclear matter in the mean field (Hartree) approximation. 
In the uniform and rotational invariant nuclear matter, the ground-state expectation value of the spatial component of the $\omega$ meson fields is zero. 
Therefore, below, we only work with the expectation value $<\omega^0>$ and write it in the symbol of $\omega$. 
We also write the expectation value $<\sigma >$ in the symbol of $\sigma$.  

The $\Sigma_{\rm s}$ and $\gamma^\mu\Sigma_\mu$ are the self-energies of the nucleon. (See Fig. 1(a).) 
Since $\Sigma^i(i=1,2,3)$ has at least one spatial component of the vector meson fields, they also become zero in the mean field approximation. 
Below, we write $\Sigma^0$ as $\Sigma_{\rm v}$. 
In the Lagrangian (\ref{eq:E1}), we have neglected the other parts of the self-energies which vanish in the mean-field approximation. 
(For example, the tensor part $\bar{\psi}[\gamma^\mu,\gamma^\nu ]\Sigma_{\mu\nu}\psi$ vanishes in the mean field approximation, since $\Sigma_{\mu\nu}$ is antisymmetric in the subscripts $\mu$ and $\nu$ and includes at least one $\omega^i$. )


Although we use the effective Lagrangian (\ref{eq:E1}) with the mean field approximation for mesonic field, our model includes the large classes of the relativistic nuclear models. 
It includes the original Walecka model,\cite{rf:Walecka,rf:Serot} the relativistic Hartree approximation (RHA),\cite{rf:Chin,rf:Serot} the nonlinear $\sigma$-$\omega$ model with $\sigma$-meson self-interactions \cite{rf:Boguta,rf:Reinhard,rf:Waldhauser,rf:Sharma,rf:Kouno1,rf:Lalazissis,rf:Iwasaki} and the $\omega$-meson self-interaction, \cite{rf:Bodmer,rf:Sugahara,rf:Kouno2} the model including $\sigma$-$\omega$ meson interaction \cite{rf:Moncada} and so forth.  
It also includes the Zimanyi and Mozkowski (ZM) model after the fermion wave function is rescaled. \cite{rf:Zimanyi} 
In the quark-meson coupling model (QMC), \cite{rf:Guichon,rf:Saito,rf:Saito2} the self-energies of the nucleon are calculated by using the bag model. \cite{rf:MIT} 

We regard the Lagrangian (\ref{eq:E1}) as an effective one in which the quantum effects of the vacuum fluctuations have been already included in the higher-order interactions. 
The $\Sigma_{\rm s}$, $\Sigma_{\rm v}$ and $U_{\rm M}$ may be able to be calculated perturbatively \cite{rf:Chin,rf:Serot} or nonperturbatively \cite{rf:Kouno4} from the bare Lagrangian which does not include the effects of the vacuum fluctuations. 
For example, in the one-loop approximation or the relativistic Hartree approximation (RHA), $U_{\rm M}$ is given by \cite{rf:Chin,rf:Serot} 
\begin{eqnarray}
U_{\rm M}(\sigma )&=&
{1\over{2}}m_\sigma^2\sigma^2-{1\over{2}}m_\omega^2\omega_\mu\omega^\mu 
-{1\over{4\pi^2}}\left\{{ (m-g_{\sigma}\sigma )^4\log{\left({m-g_\sigma\sigma\over{m}}\right)} }\right.
\nonumber\\
&&\left.+g_\sigma m^3\sigma-{7\over{2}}g_\sigma^2m^2\sigma^2+{13\over{3}}g_\sigma^3m\sigma^3-{25\over{12}}
g_\sigma^4\sigma^4\right\}. 
\label{eq:A20a}
\end{eqnarray}
However, there are much ambiguity even in the 1-loop approximations \cite{rf:Heide,rf:Cohen2,rf:Kouno3} and the higher order calculations are much more difficult, we treat $\Sigma_{\rm s}$, $\Sigma_{\rm v}$ and $U_{\rm M}$ as phenomenological inputs rather than to calculate them from the bare Lagrangian. 
Therefore, in principle, there are a limitless number of parameters, namely, effective couplings in these quantities. 

We also regard the density effects beyond the Hartree approximation 
are included in the explicit density-dependence through $S$ and $V$. 
So the effective Lagrangian (\ref{eq:E1}) also include the density dependent field theory, at least in its simplest form. 
One important thing concerned the explicit density dependence is that the mesonic potential $U_{\rm M}$ is not independent of the nucleon self-energies $\Sigma_{\rm s}$ and $\Sigma_{\rm v}$. 
As is in the original density dependent hadron field theory, \cite{rf:Fuchs} they are related with each other through the thermodynamical equation
\begin{eqnarray}
{d\epsilon (\rho_{\rm B},\phi_i)\over{d\rho_{\rm B}}}=\mu, 
\label{eq:E2}
\end{eqnarray}
where $\epsilon$ and $\mu$ are the energy density and the baryonic chemical potential in the system, and $\phi_i =\sigma$, $\omega$. 
The $\rho_{\rm B}$ in $\epsilon (\rho_{\rm B}, \phi_i)$ represents explicit 
density dependence through the fermi momentum $k_{\rm F}$. 


Using the equations of motion 
\begin{eqnarray}
{\partial \epsilon (\rho_{\rm B},\phi_i)\over{\partial \phi_i}}=0, 
\label{eq:E3}
\end{eqnarray}
we obtain 
\begin{eqnarray}
{d\epsilon (\rho_{\rm B},\phi_i)\over{d\rho_{\rm B}}}
&=&{\partial\epsilon (\rho_{\rm B},\phi_i)\over{\partial \rho_{\rm B}}}
+\sum_{i}{d\phi_i\over{d\rho_{\rm B}}}{\partial \epsilon\over{\partial\phi_i}}
\nonumber\\
&=&{\partial \epsilon (\rho_{\rm B},\phi_i)\over{\partial \rho_{\rm B}}}=\mu. 
\label{eq:E4}
\end{eqnarray}
We remark that this equation is nothing but the evolution equation which determine the explicit density dependence of effective masses and couplings which is defined at finite density. 
The Legendre transformed version of this equation, namely, 
\begin{eqnarray}
{\partial P(\mu, \phi_i)\over{\partial \mu}}=\rho_{\rm B}
\label{eq:E5}
\end{eqnarray}
is the chemical potential flow equation proposed by 
Berges, Jungnickel and Wetterich. \cite{rf:Berges} 


Using the mean field approximation, we can represent the energy density and the pressure in term of the fermi momentum $k_{\rm F}$ as follows. 
\begin{eqnarray}
\epsilon (k_{\rm F},\sigma,\omega,S,V)&=&\epsilon_{\rm B}(k_{\rm F},\sigma,\omega,S,V)+U_{\rm M}(\sigma,\omega,S,V)-\Sigma_{\rm v}(\sigma,\omega,S,V)\rho_{\rm B}(k_{\rm F}); 
\nonumber\\
\epsilon_B(k_{\rm F},m^*)
&=&{1\over{6\pi^2}}\left\{ E_{\rm F}^* k_{\rm F}(3k_{\rm F}^2+{3\over{2}}{m^*}^2)-{3\over{2}}{m^*}^4\log{\left({E_{\rm F}^*+k_{\rm F}\over{m^*}}\right)}\right\}; 
\nonumber\\
m^*&=&m+\Sigma_{\rm s}(\sigma,\omega,S,V),~~~~~~E_{\rm F}^*=\sqrt{k_{\rm F}^2+{m^*}^2} 
\label{eq:E6}
\end{eqnarray}
and 
\begin{eqnarray}
P&=&P_{\rm B}(k_{\rm F},m^*)-U_{\rm M}(\rho_{\rm B},\sigma,\omega,S,V); 
\nonumber\\
P_B(\rho_{\rm B},m^*)&=&{1\over{6\pi^2}}\left\{ E_{\rm F}^* k_{\rm F}(k_{\rm F}^2-{3\over{2}}{m^*}^2)+{3\over{2}}{m^*}^4\log{\left({E_{\rm F}^*+k_{\rm F}\over{m^*}}\right)}\right\}. 
\label{eq:E7}
\end{eqnarray}
Equations (\ref{eq:E6}), (\ref{eq:E7}) and the thermodynamical identity 
$\mu =(\epsilon +P)/\rho_{\rm B}$ yields 
\begin{eqnarray}
\mu =E_{\rm F}^*-\Sigma_{\rm v}.
\label{eq:E8}
\end{eqnarray}


According to Fuchs, Lenske and Wolter, \cite{rf:Fuchs} we put Eqs. (\ref{eq:E6}) and (\ref{eq:E8}) into (\ref{eq:E4}). 
We obtain 
\begin{eqnarray}
{d\epsilon (k_{\rm F},\phi_i)\over{d\rho_{\rm B}}}
&=&{dk_{\rm F}\over{d\rho_{\rm B}}}{\partial\epsilon (k_{\rm F},\phi_i)\over{\partial k_{\rm F}}}
+\sum_{i}{d\phi_i\over{d\rho_{\rm B}}}{\partial \epsilon\over{\partial\phi_i}}
+{d S\over{d \rho_{\rm B}}}{\partial \epsilon\over{\partial S}}
+{d V\over{d \rho_{\rm B}}}{\partial \epsilon\over{\partial V}}
\nonumber\\
&=&
E_{\rm F}^*-\Sigma_{\rm v}
+{d S\over{d\rho_{\rm B}}}{\partial \epsilon\over{\partial S}}
+{d V\over{d \rho_{\rm B}}}{\partial \epsilon\over{\partial V}}
\nonumber\\
&=&E_{\rm F}^*-\Sigma_{\rm v}
+{d\rho_{\rm s}\over{d\rho_{\rm B}}}{\partial \epsilon\over{\partial S}}
+{\partial \epsilon\over{\partial V}}
=E_{\rm F}^*-\Sigma_{\rm v} 
\label{eq:E9}
\end{eqnarray}

Therefore, we obtain 
\begin{eqnarray}
{d\rho_{\rm s}\over{d\rho_{\rm B}}}{\partial \epsilon\over{\partial S}}
+{\partial \epsilon\over{\partial V}}=0. 
\label{eq:E10}
\end{eqnarray}
We determine the forms of $\Sigma_{\rm s}$, $\Sigma_{\rm v}$ and $U_{\rm M}$ such that this equation has the solution $S=\rho_{\rm s}$ and $V=\rho_{\rm B}$. 

It is difficult to find the general forms of $\Sigma_{\rm s}$, $\Sigma_{\rm v}$ and $U_{\rm M}$ for Eq. (\ref{eq:E10}). 
However, if we assume 
\begin{eqnarray}
{\partial \epsilon\over{\partial S}}
={\partial \Sigma_{\rm s}\over{\partial S}}\rho_{\rm s}
-{\partial \Sigma_{\rm v}\over{\partial S}}\rho_{\rm B}
+{\partial U_{\rm M}\over{\partial S}}
=0, 
\label{eq:E11}
\end{eqnarray}
equations (\ref{eq:E10}) and (\ref{eq:E11}) give 
\begin{eqnarray}
{\partial \epsilon\over{\partial V}}
={\partial \Sigma_{\rm s}\over{\partial V}}\rho_{\rm s}
-{\partial \Sigma_{\rm v}\over{\partial V}}\rho_{\rm B}
+{\partial U_{\rm M}\over{\partial V}}
=0. 
\label{eq:E12}
\end{eqnarray}


The coupled equations (\ref{eq:E11}) and (\ref{eq:E12}) 
have a analytic solution 
\begin{eqnarray}
\Sigma_{\rm s}=\sum_{m,n}c_{m,n}(\sigma,\omega)S^mV^n,~~~~~~
\Sigma_{\rm v}=\sum_{m,n}d_{m,n}(\sigma,\omega)S^mV^n
\label{eq:E12a} 
\end{eqnarray}
and 
\begin{eqnarray}
U_{\rm M}=\sum_{m,n}u_{m,n}(\sigma,\omega)S^mV^n, 
\label{eq:E13}
\end{eqnarray}
if the conditions 
\begin{eqnarray}
c_{m,n}&=&-{m+1\over{m+n}}u_{m+1,n}~~(m\geq 0,n\geq 0,m+n\neq 0),~~~
\nonumber\\
d_{m,n}&=&{n+1\over{m+n}}u_{m,n+1}~~(m\geq 0,n\geq 0,m+n\neq 0),~~~
\nonumber\\
u_{0,1}&=&0,~~~u_{1,0}=0
\label{eq:E14}
\end{eqnarray}
and
\begin{eqnarray}
\det\left|
\left[
\begin{array}{cc}
{\partial \Sigma_{\rm s}\over{\partial S}} &
-{\partial \Sigma_{\rm v}\over{\partial S}} \\
             &                              \\ 
-{\partial \Sigma_{\rm s}\over{\partial V}} &
{\partial \Sigma_{\rm v}\over{\partial V}}
\end{array}
\right]\right| \neq 0
\label{eq:E15} 
\end{eqnarray}
are satisfied. 
Because of Eqs. (\ref{eq:E11}) and (\ref{eq:E12}), we can regard 
$S$ and $V$ as auxiliary boson fields.  


We summarize the other basic quantities we will use later. 
In the mean field approximation, the nucleon propagator $G^{\rm N}(k)$ is given by \cite{rf:Walecka, rf:Chin}
\begin{equation}
G_{\rm N}(k)=G^{\rm F}_{\rm N}(k)+G_{\rm N}^{\rm D}(k)
\label{eq:E16}
\end{equation}
with the Feynman part 
\begin{equation}
G^{\rm F}_{\rm N}(k)=(\gamma^\mu k_\mu^*+m^*){-1\over{-{k^*}^2+{m^*}^2-i\epsilon}}
\label{eq:E17}
\end{equation}
and the density part 
\begin{equation}
G^{\rm D}_{\rm N}(k)=(\gamma^\mu k_\mu^*+m^*){i\pi\over{E_k^*}}\delta ({k^*}^0-E_k^*)\theta (k_{\rm FN}-\vert {\bf k}\vert ), 
\label{eq:E18}
\end{equation}
where ${k^*}^\mu=(k^0+\Sigma_{\rm v},{\bf k})$. 
Since all of the vacuum fluctuations effects have been already included in the effective Lagrangian (\ref{eq:E1}), we use only the density part $G^{\rm D}_{\rm N}(k)$ to evaluate the density effects. 

Using the propagator $G^{\rm D}_{\rm N}(k)$, we get 
the baryon density 
\begin{eqnarray}
\rho_B 
&=& <\bar{\psi}\gamma^0\psi >
=-i\int{d^4k\over{(2\pi)^4}}{\rm Tr}[\gamma^0G^{\rm D}_{\rm N}(k)]
={2\over{3\pi^2}}k_{{\rm F}}^3. 
\label{eq:19}
\end{eqnarray}

The scalar density is also given by 
\begin{eqnarray}
\rho_{\rm s}(k_{\rm F},m^*)&=&<\bar{\psi}\psi >
=-i\int{d^4k\over{(2\pi)^4}}{\rm Tr}[G^{\rm D}_N(k)]
\nonumber\\
&=&{1\over{\pi^2}}m^*\left\{k_{\rm F}E_{\rm F}^*-{m^*}^2\ln{\left({k_{\rm F}+E_{\rm F}^*\over{m^*}}\right)}\right\}. 
\label{eq:E20}
\end{eqnarray}


The equation of motion for $\sigma$-meson is given by 
\begin{equation}
{\partial \epsilon(\rho_{\rm B},\sigma,\omega,S,V)\over{\partial \sigma}}=0. 
\label{eq:E20a}
\end{equation}
Putting (\ref{eq:E6}) into (\ref{eq:E20a}), we get 
\begin{eqnarray}
&&{\partial \epsilon_{\rm B}(k_{\rm F},m^*(\sigma,\omega,S,V))\over{\partial \sigma}}
-{\partial \Sigma_{\rm v}(\sigma,\omega,S,V)\over{\partial \sigma}}\rho_{\rm B}
\nonumber\\
&&
+{\partial U_{\rm M}(\sigma,\omega,S,V)\over{\partial \sigma}}
\nonumber\\
&=&{\partial \Sigma_{\rm s}(\sigma,\omega,S,V)\over{\partial \sigma}}\rho_{\rm s}
-{\partial \Sigma_{\rm v}(\sigma,\omega,S,V)\over{\partial \sigma}}\rho_{\rm B}
+{\partial U_{\rm M}(\sigma,\omega,S,V)\over{\partial \sigma}}
\nonumber\\
&=&-g_{\rm s\sigma}^*\rho_{\rm s}+g_{\rm v\sigma}^*\rho_{\rm B}
+{\partial U_{\rm M}(\sigma,\omega,S,V)\over{\partial \sigma}}=0. 
\label{eq:E21}
\end{eqnarray}
In eq. (\ref{eq:E21}), we have used the relation 
\begin{equation}
{\partial \epsilon_{\rm B}(k_{\rm F},m^*)\over{\partial m^*}}=\rho_{\rm s}
\label{eq:E22}
\end{equation}
and have defined the effective couplings for the three-point meson-nucleon interaction as 
\begin{equation}
g_{\rm s\sigma}^*\equiv 
-{\partial \Sigma_{\rm s}(\sigma,\omega,S,V )\over{\partial \sigma}}
~~~{\rm and}~~~
g_{\rm v\sigma}^*\equiv 
-{\partial \Sigma_{\rm v}(\sigma,\omega,S,V)\over{\partial \sigma}}
\label{eq:E23}
\end{equation}


Note that the differentiating the nucleon self-energies with respect to the meson-field expectation value yields the effective couplings of the meson-nucleon interaction. 
One external line of the meson can be attached at the point where one meson mean field have been removed by the differentiation. 
(See Figs. 1(a) and (b).) 
In general, the effective action is a generating functional of one-particle-irreducible correlation functions, namely, effective masses and effective couplings. \cite{rf:Peskin} 

Similarly, the equation of motion for $\omega$-meson is given by 
\begin{eqnarray}
&&-g_{\rm s\omega}^*\rho_{\rm s}+g_{\rm v\omega}^*\rho_{\rm B}
+{\partial U_{\rm M}(\sigma,\omega,S,V)\over{\partial \omega}}=0, 
\label{eq:E24}
\end{eqnarray}
where we have also defined the effective couplings 
\begin{equation}
g_{\rm s\omega}^*\equiv 
-{\partial \Sigma_{\rm s}(\sigma,\omega,S,V)\over{\partial \omega}}
~~~{\rm and}~~~
g_{\rm v\omega}^*\equiv 
-{\partial \Sigma_{\rm v}(\sigma,\omega,S,V )\over{\partial \omega}}. 
\label{eq:E25}
\end{equation}
The diagrammatic description for Eqs. (\ref{eq:E21}) and (\ref{eq:E24}) is shown in Fig. 1(c). 
Although there are a limitless number of parameters in $\Sigma_{\rm s}$ and $\Sigma_{\rm v}$, only four effective couplings appear for the meson-nucleon interaction in Eqs. (\ref{eq:E21}) and (\ref{eq:E24}). 
If we put $g_{\rm v\sigma}^*=g_{\rm s\omega}^*=0$ and approximate $g_{\rm s\sigma}^*$ and $g_{\rm v\omega}^*$ as constants which are determined at the normal density, we have familiar equations of motion which are used in the original Walecka model, \cite{rf:Walecka,rf:Serot} the RHA calculation \cite{rf:Chin} and the nonlinear $\sigma$-$\omega$ model. \cite{rf:Boguta,rf:Reinhard,rf:Waldhauser,rf:Bodmer,rf:Sharma,rf:Sugahara,rf:Moncada,rf:Kouno1,rf:Kouno2,rf:Lalazissis,rf:Iwasaki}


Since the effective potential is a generating function of one-particle-irreducible correlation functions with vanishing external momentum, \cite{rf:Peskin} 
the second derivatives with respect to the meson mean fields yield the effective meson masses. 
In fact, differentiating the l.h.s. of the equations of motion (\ref{eq:E21}) with respect to the meson-field expectation values, we get the equation for the meson self-energies \cite{rf:Walecka,rf:Chin,rf:Tuchitani} with vanishing external momentum, namely, 
\begin{eqnarray}
&&{\partial^2 \epsilon (\rho_{\rm B},\sigma,\omega,S,V)\over{\partial \sigma^2}}
\nonumber\\
&=&-i2{g_{\rm s\sigma}^*}^2\int{d^4k\over{(2\pi)^4}}
{\rm Tr}[G^{\rm F}_{\rm N}(k)G^{\rm D}_{\rm N}(k)+G^{\rm D}_{\rm N}(k)G^{\rm F}_{\rm p}(k)+G^{\rm D}_{\rm N}(k)G^{\rm D}_{\rm N}(k)]
\nonumber\\
&&+i2g_{\rm s\sigma\sigma}^*\int{d^4k\over{(2\pi)^4}}{\rm Tr}[G^{\rm D}_{\rm N}(k)]
-ig_{\rm v\sigma\sigma}^*\int{d^4k\over{(2\pi)^4}}{\rm Tr}[\gamma^0G^{\rm D}_{\rm N}(k)]
\nonumber\\
&&+{\partial^2 U_{\rm M}(\sigma,\omega,S,V )\over{\partial \sigma^2}}
\nonumber\\ 
&\equiv &{g_{\rm s\sigma}^*}^2\Pi-g_{\rm s\sigma\sigma}^*\rho_{\rm s}
+g_{\rm v\sigma\sigma}^*\rho_{\rm B}
+{\partial^2 U_{\rm M}(\sigma,\omega,S,V)\over{\partial \sigma^2}}. 
\label{eq:E31}
\end{eqnarray}
Similarly, in general, we get
\begin{eqnarray}
{\partial^2 \epsilon (\rho_{\rm B},\sigma,\omega,S,V)\over{\partial \phi_i\partial \phi_j}}
&= &g_{{\rm s}{\phi_i}}^*g_{{\rm s}{\phi_j}}^*\Pi-g_{{\rm s}{\phi_i}{\phi_j}}^*\rho_{\rm s}
+g_{{\rm v}{\phi_i}{\phi_j}}^*\rho_{\rm B}
\nonumber\\
&+&{\partial^2 U_{\rm M}(\sigma,\omega,S,V)\over{\partial\phi_i\partial \phi_j}},~~~~~~(i,j=1,2)
\label{eq:E32}
\end{eqnarray}
where $\phi_1=\sigma$ and $\phi_2=\omega$, and the effective coupling $g_{{\rm f}\phi_i\phi_j}^*$ (f=s,v) is defined by
\begin{equation}
g_{{\rm f}{\phi_i}{\phi_j}}^*\equiv 
-{\partial^2 \Sigma_{\rm f}(\sigma,\omega,S,V )\over{\partial\phi_i\partial\phi_j}}. 
\nonumber
\label{eq:E33}
\end{equation}
The diagrammatic descriptions for Eqs. (\ref{eq:E31}) and (\ref{eq:E32}) are shown in  Fig. 1(d). 
If the mixing part ${\partial^2 \epsilon\over{\partial \sigma\partial\omega}}$ can be neglected, ${\partial^2 \epsilon\over{\partial \sigma^2}}$ and $-{\partial^2 \epsilon\over{\partial \omega^2}}$ are the effective meson masses which are defined at the zero external momentum. 


\section{Incompressibility of nuclear matter}

In this section, we derive the simple relation among the effective masses, effective couplings and incompressibility. 

At first we study the density evolution of the expectation values $\sigma$ and $\omega$. 
We differentiate the equations of motion 
(\ref{eq:E21}) for $\sigma$-meson with respect to $\rho_{\rm B}$. 
We obtain 
\begin{eqnarray}
&&{\partial\over{\partial \rho_{\rm B}}}{\partial \epsilon (\rho_{\rm B},\phi_i)\over{\partial \sigma}}
+\sum_{j}{d\phi_j\over{d\rho_{\rm B}}}{\partial^2\epsilon (\rho_{\rm B},\phi_i )\over{\partial \phi_j\partial \sigma}}=0
\label{eq:E34}
\end{eqnarray}
where we have redefined $\phi_i=\sigma,\omega,S,V$. 
We also get 
\begin{eqnarray}
{\partial\over{\partial \rho_{\rm B}}}{\partial \epsilon (\rho_{\rm B},\phi_i)\over{\partial\phi_j }}
&=&
{\partial\over{\partial \rho_{\rm B}}}{\partial \epsilon_{\rm B} (k_{\rm F}(\rho_{\rm B}), m^*(\phi_i))\over{\partial\sigma }}
\nonumber\\
&-&
{\partial\over{\partial \rho_{\rm B}}}\left({\partial \Sigma_{\rm v}(\phi_i )\over{\partial \sigma}}\rho_{\rm B}\right)
\nonumber\\
&=&
{\partial\over{\partial \rho_{\rm B}}}\left({\partial \Sigma_{\rm s}(\phi_i)\over{\partial \sigma}}{\partial \epsilon_{\rm B} (k_{\rm F},m^*)\over{\partial m^* }}\right)
\nonumber\\
&-&{\partial \Sigma_{\rm v}(\phi_i )\over{\partial \sigma}}
\nonumber\\
&=&
-g_{\rm s\sigma}^*{\partial \rho_{\rm s}(k_{\rm F}, m^*)\over{\partial \rho_{\rm B}}}+g_{\rm v\sigma}^* =-g_{\rm s\sigma}^*{\hat{\gamma}}^{-1}+g_{\rm v\sigma}^*, 
\nonumber\\
&&\label{eq:E38}
\end{eqnarray}
where ${\hat{\gamma}}$ is an effective gamma factor which is defined as 
\begin{equation}
{\hat{\gamma}}^{-1}\equiv{\partial \rho_{\rm s}(k_{\rm F}(\rho_{\rm B}),m^*)\over{\partial \rho_{\rm B}}}
={m^*\over{E_{\rm F}^*}}
\label{eq:E39}
\end{equation}


Therefore, we obtain
\begin{eqnarray}
\sum_{j}{d\phi_j\over{d\rho_{\rm B}}}{\partial^2\epsilon (\rho_{\rm B},\phi_i )\over{\partial \phi_j\partial \sigma}}
&=&g_{\rm s\sigma}^*{\hat{\gamma}}^{-1}-g_{\rm v\sigma}^*. 
\label{eq:E40a}
\end{eqnarray}
Similar equations can be obtained for the other fields, namely, $\omega$, $S$ and $V$. 
Combining these equations, we obtain 
\begin{equation}
{M^*}^2{d\over{d\rho_{\rm B}}}{\bf \Phi} =-{\hat{\bf g}}, 
\label{eq:E40}
\end{equation}
where 
\begin{equation}
{\hat{\bf g}}\equiv -{\hat{\gamma}}^{-1}{\bf g}_{\rm s}^*+{\bf g}_{\rm v}^*
;~~~~~ 
{\bf g}_{\rm s}^*\equiv 
\left[
\begin{array}{c}
g_{\rm s\sigma}^* \\
g_{\rm s\omega}^* \\
g_{\rm sS}^*   \\
g_{\rm sV}^*
\end{array}
\right], 
~~~~~
{\bf g}_{\rm v}^*\equiv 
\left[
\begin{array}{c}
g_{\rm v\sigma}^* \\
g_{\rm v\omega}^* \\
g_{\rm vS}^*   \\
g_{\rm vV}^*
\end{array}
\right],  
\label{eq:E41}
\end{equation}
\begin{equation}
{\bf \Phi}
\equiv 
\left[
\begin{array}{c}
\sigma \\
\omega \\
S      \\
V
\end{array}
\right], 
\label{eq:E42}
\end{equation}
and 
\begin{equation}
{M^*}^2
\equiv \left[
\begin{array}{cccc}
{m_{\sigma}^*}^2 & {m_{\sigma\omega}^*}^2 & {m_{\sigma S}^*}^2 
& {m_{\sigma V}^*}^2 \\
{m_{\sigma\omega}^*}^2 & -{m_{\omega}^*}^2 & {m_{\omega S}^*}^2 
& {m_{\omega V}^*}^2 \\
{m_{\sigma S}^*}^2 & {m_{\omega S}^*}^2 & {m_{S}^*}^2
& {m_{S V}^*}^2 \\
{m_{\sigma V}^*}^2 & {m_{\omega V}^*}^2 & {m_{S V}^*}^2
& -{m_{V}^*}^2 \\
\end{array}
\right]
\equiv 
\left[
\begin{array}{cccc}
{\partial^2 \epsilon\over{\partial \sigma^2}} & {\partial^2 \epsilon\over{\partial \sigma \partial\omega}} & {\partial^2 \epsilon\over{\partial\sigma \partial S}} & {\partial^2\epsilon\over{\partial \sigma \partial V}} \\
{\partial^2 \epsilon\over{\partial\sigma\partial\omega}} & {\partial^2 \epsilon\over{\partial \omega^2}} & {\partial^2 \epsilon\over{\partial \omega \partial S}} & {\partial^2\epsilon\over{\partial \omega \partial V}} \\
{\partial^2 \epsilon\over{\partial\sigma \partial S}} & {\partial^2 \epsilon\over{\partial\omega \partial S}} & {\partial^2 \epsilon\over{\partial S^2}}
& {\partial^2\epsilon\over{\partial S \partial V}} \\
{\partial^2 \epsilon\over{\partial\sigma \partial V}} & {\partial^2 \epsilon\over{\partial\omega \partial V}} & {\partial^2 \epsilon\over{\partial S\partial V}}& {\partial^2\epsilon\over{\partial V^2}} \\
\end{array}
\right] .
\label{eq:E43}
\end{equation}
If $\det{{M^*}^2}\neq 0$, 
Eq. (\ref{eq:E40}) can be transformed as 
\begin{equation}
{d{\bf \Phi}\over{d\rho_{\rm B}}} =-({M^*}^2)^{-1}{\hat{\bf g}}
\label{eq:E45}
\end{equation}


The incompressibility $K$ is defined by  
\begin{equation}
K\equiv 
9\rho_{\rm B0}^2\left.{d^2 (\epsilon /\rho_{\rm B} ) \over{d \rho_{\rm B}^2}}\right|_{\scriptstyle\rho_{\rm B} =\rho_{\rm B0}\atop\scriptstyle \rho_3=0}
=
9\left.{d P\over{d \rho_{\rm B}}}\right|_{\scriptstyle\rho_{\rm B} =\rho_{\rm B0}\atop\scriptstyle \rho_3=0}
=
9\rho_{\rm B0}\left.{d \mu\over{d \rho_{\rm B}}}\right|_{\scriptstyle\rho_{\rm B} =\rho_{\rm B0}\atop\scriptstyle \rho_3=0}, 
\label{eq:E64}
\end{equation}
where $\rho_{\rm B0}$ is the normal baryon density. 
Using 
$\mu =E_{\rm F}^*-\Sigma_{\rm v}=\sqrt{k_{\rm F}^2+{m^*}^2}-\Sigma_{\rm v}$ and Eq. (\ref{eq:E45}),  
we get 
\begin{eqnarray}
{1\over{\rho}}{dP\over{d\rho_{\rm B}}}&=&{d\mu\over{d\rho_{\rm B}}}
=
{dk_{\rm F}\over{d\rho}}
{k_{\rm F}\over{E_{\rm F}^*}}
\nonumber\\
&+&{m^*\over{E_{\rm F}^*}}\left(\sum_{j}
 {\partial \Sigma_{\rm s}(\phi_i)\over{\partial \phi_j}}{d\phi_j\over{d\rho_{\rm B}}}\right)
-\left(\sum_{j}
 {\partial \Sigma_{\rm v}(\phi_i)\over{\partial \phi_j}}{d\phi_j\over{d\rho_{\rm B}}}\right)
\nonumber\\
&=&{k_{\rm F}^2\over{3\rho_{\rm B} E_{\rm F}^*}}+^t{\hat{\bf g}}
{d{\bf \Phi}\over{d\rho_{\rm B}}}
={k_{\rm F}^2\over{3\rho_{\rm B}E_{\rm F}^*}}-^t{\hat{\bf g}} ({M^*}^2)^{-1}{\hat{\bf g}}. 
\label{eq:E65}
\end{eqnarray}
Therefore, we get the relation among the effective masses, the effective couplings and the incompressibility. 
\begin{equation}
K=9\rho_{\rm B}\left.\left({k_{\rm F}^2\over{3\rho_{\rm B} E_{\rm F}^*}}-^t{\hat{\bf g}}({M^*}^2)^{-1}{\hat{\bf g}}\right)\right|_
{\scriptstyle\rho_{\rm B} =\rho_{\rm B0} \atop\scriptstyle\rho_3=0}. 
\label{eq:E66}
\end{equation}
In particular, if $({M^*}^2)^{-1}$ is diagonal and there are no explicit density dependence, 
\begin{equation}
-^t{\hat{\bf g}}({M^*}^2)^{-1}{\hat{\bf g}}
={{\hat{g}_{\omega}}^2\over{{m_{\omega}^*}^2}}
-{{\hat{g}_{\sigma}}^2\over{{m_{\sigma}^*}^2}}
={(g_{\rm v\omega}^*-\hat{\gamma}^{-1}g_{\rm s\omega}^*)^2\over{{m_{\omega}^*}^2}}
-{(g_{\rm v\sigma}^*-g_{\rm s\sigma}\hat{\gamma}^{-1})^2\over{{m_{\sigma}^*}^2}}.\label{eq:E67} 
\end{equation}
Therefore, this quantity represents the difference between the strengths of the effective repulsive force and the effective attractive force. 
From Eqs. (\ref{eq:E65}) and (\ref{eq:E67}), it is easily seen that the reduction of the effective $\omega$-meson mass and/or the enhancement of the effective $\omega$-nucleon coupling make the EOS stiffer. 
We remark that the effective $\sigma$-meson mass $m_\sigma^*$ in Eq. (\ref{eq:E67}) includes the contribution in the ordinary random phase approximation (RPA) for the meson self-energy, namely $\Pi$ in Eq. (\ref{eq:E31}). 
Therefore, Eqs. (\ref{eq:E66}) and (\ref{eq:E67}) are the generalization of the earlier result. \cite{rf:Matsui} 

From (\ref{eq:E66}), we see that 
the value of $-^t\hat{\bf g}({M^*}^2)^{-1}\hat{\bf g}$ at the normal density 
is determined if the values of $K$ and $m_0^*$ (subscript "0" means that the quantities with it are defined at the normal density $\rho_{\rm B0}$) are given. 
In Fig. 2, we show the relation among $m_0^*$, $K$ and 
$-^t\hat{\bf g}({{M}^*}^2)^{-1}\hat{\bf g}$. 
For the well-known parameter sets, the values of $-^t\hat{\bf g}({{M}^*}^2)^{-1}\hat{\bf g}$ almost vanish. \cite{rf:Tuchitani} 
The effective attractive and repulsive forces almost cancel each other at the normal density. 


Because of the identity $V=\rho_{\rm B}$, we can show that 
\begin{eqnarray}
{m_V^*}^2=g_{\rm vV}^*-{m^*\over{E_{\rm F}^*}}g_{\rm sV}^*=\hat{g}_V. 
\label{eq:E67a}
\end{eqnarray}
Similarly, if 
\begin{eqnarray}
\omega = {g_\omega\over{m_\omega^2}}\rho_{\rm B}, 
\label{eq:E67b}
\end{eqnarray}
we obtain 
\begin{eqnarray}
{{m_\omega^*}^2\over{{m_\omega}^2}}= {g_{\rm v\omega}^*-{m^*\over{E_{\rm F}^*}}g_{\rm s\omega}^*\over{g_\omega}}={\hat{g}_\omega \over{g_\omega}}.   
\label{eq:E67c}
\end{eqnarray}
In the original Walecka model, \cite{rf:Walecka} Eq. (\ref{eq:E67b}) is exact and Eq. (\ref{eq:E67c}) is trivial, since $m_\omega^*=m_\omega$, $g_{\rm v\omega}^*=g_\omega$ and $g_{\rm s\omega}^*=0$. 
Furthermore, in a viewpoint of the quark physics, it may be natural that the $\omega$-meson mean field is proportional to the baryon density. \cite{rf:Kunihiro} 
There are non-trivial cases which satisfy the relation (\ref{eq:E67c}). 
In such a case, because of the relation (\ref{eq:E67c}), the effective $\omega$-nucleon coupling automatically decreases if the effective $\omega$-meson mass decreases. 
Putting (\ref{eq:E67c}) into (\ref{eq:E67}), we obtain 
\begin{equation}
-^t{\hat{\bf g}}({M^*}^2)^{-1}{\hat{\bf g}}
={{\hat{g}_{\omega}}^2\over{{m_{\omega}^*}^2}}
-{{\hat{g}_{\sigma}}^2\over{{m_{\sigma}^*}^2}}
={{\hat{g}_{\omega}}g_\omega\over{{m_{\omega}}^2}}
-{{\hat{g}_{\sigma}}^2\over{{m_{\sigma}^*}^2}}.
\label{eq:E67e} 
\end{equation}
If the effective $\omega$-meson mass becomes smaller, $\hat{g}_\omega$ also becomes smaller according to Eq. (\ref{eq:E67c}). 
Therefore, the EOS becomes softer according to Eq. (\ref{eq:E67e}). 
In the next section, we will show examples of such EOS. 


\section{Effective hadron masses, effective coupling and equation of state}

In this section, we numerically investigate relation among effective hadron masses, effective couplings and equation of states (EOS). 
Below we assume that
\begin{eqnarray}
\Sigma_{\rm s}(\sigma )&=&-g_\sigma\sigma -g_{\sigma S}\sigma S, 
\label{eq:E69}\\
\Sigma_{\rm v} (\omega )&=&-g_\omega\omega -g_{\omega 3}\omega^3-g_{\omega V2}\omega V^2-g_{\omega 2 V}\omega^2V, 
\label{eq:E70}
\end{eqnarray}
and 
\begin{eqnarray}
U_{\rm M}(\sigma ,\omega )&=&{1\over{2}}m_{\sigma}^2\sigma^2+{1\over{3}}c_{\sigma 3}\sigma^3+{1\over{4}}c_{\sigma 4}\sigma^4
-
{1\over{2}}m_{\omega}^2\omega^2-{1\over{4}}c_{\omega 4}\omega^4
\nonumber\\
&+&
{1\over{2}}c_{\sigma 2\omega 2}\sigma^2\omega^2+{1\over{2}}g_{\sigma S}\sigma S^2-{2\over{3}}g_{\omega V3}\omega V^3-{1\over{2}}g_{\omega 2 V}\omega^2V^2
\nonumber\\
&&
\label{eq:E72}
\end{eqnarray}
Under these assumptions, 
the effective meson-nucleon couplings and the effective meson masses are given by the following equations. 
\begin{eqnarray}
{\bf g}_{\rm s}^*
=
\left[
\begin{array}{c}
g_{\rm s\sigma}^* \\
                  \\
g_{\rm s\omega}^* 
\end{array}
\right]
=
\left[
\begin{array}{c}
g_\sigma +g_{\sigma S}S \\
                        \\
0 
\end{array}
\right]. 
\label{eq:E73}
\end{eqnarray}
\begin{eqnarray}
{\bf g}_{\rm v}^*
=\left[
\begin{array}{c}
g_{\rm v\sigma}^* \\
                  \\
g_{\rm v\omega}^* 
\end{array}
\right]
=\left[
\begin{array}{c}
0 \\
  \\
g_\omega+3g_{\omega 3}\omega^2+g_{\omega V2}V^2+2g_{\omega 2 V}\omega V
\end{array}
\right]. 
\label{eq:E74}
\end{eqnarray}
\begin{eqnarray}
{m^*_\sigma}^2&=&{m_\sigma}^2+2c_{\sigma 3}\sigma+3c_{\sigma 4}\sigma^2
+c_{\sigma 2\omega 2}\omega^2
\nonumber\\
&+&{{g_{\rm s\sigma}^*}^2\over{\pi^2}}\left\{k_FE_F^*+2{k_F{m^*}^2\over{{E_F^*}}}-3{m^*}^2\log{\left({k_F+E_F^*\over{m^*}}\right)}\right\}. 
\label{eq:E76}
\\
{m^*_\omega}^2&=&{m_\omega}^2+3c_{\omega 4}\omega^2
-6g_{\omega 3}\omega\rho_{\rm B}-c_{\sigma 2\omega 2}\sigma^2
-g_{\omega 2 V}V\rho_{\rm B} 
\label{eq:E77}
\end{eqnarray}
The non-diagonal elements of the matrix $M^*$ are given by following equations. 
\begin{equation}
{m_{\sigma\omega}^*}^2=2c_{\sigma 2\omega 2}\sigma\omega. 
\label{eq:E79}
\end{equation}

At fast, we examine the effects of the higher-order interaction without 
explicit density dependence. 
For numerical calculations, 
we use the NL3 parameter set \cite{rf:Lalazissis} as a basic one, in which 
all couplings vanish except for $g_\sigma$, $g_\omega$, $c_{\sigma 3}$ and 
$c_{\sigma 4}$. 
In other sets, we add one (or two) higher order coupling(s) to $\Sigma$ or $U_{\rm M}$, and modify $g_\sigma$, $g_\omega$, $c_{\sigma 3}$ and $c_{\sigma 4}$ to 
reproduce the basic physical properties in the NL3, namely, $\rho _0=0.148$fm$^{-3}$, $a_1=16.299$MeV, $m^*_0=0.60m$ and $K=271.76$MeV. 
The parameter sets are summarized in Table 1. 


Since we concentrate ourselves about the effective $\omega$-meson mass reduction, we consider following three cases. 
In set B, the $\omega$-meson self-interaction is added with negative sign. 
From (\ref{eq:E77}), it is easily seen that this contribution makes the effective $\omega$-meson mass smaller. 
In set D, the $\omega$-$\omega$-$\omega$-nucleon interaction is added with positive sign. 
From (\ref{eq:E74}) and (\ref{eq:E77}), it is easily seen that this contribution makes the effective $\omega$-nucleon coupling larger and the effective $\omega$-meson mass smaller. 
In set K, the $\sigma$-$\omega$ mixing interaction is added with positive sign. 
From (\ref{eq:E76}) and (\ref{eq:E77}),it is easily seen that this contribution makes the effective $\sigma$-meson mass larger and the effective $\omega$-meson mass smaller. 

In parameter sets B and D, we determine the higher order interaction to satisfy ${m_\omega^*}^2/m_\omega^2=0.9$. 
In parameter sets K, we determine the higher order interaction to satisfy ${m_\omega^*}^2/m_\omega^2=0.95$. 
(We could not find the parameter set which satisfies ${m_\omega^*}^2/m_\omega^2=0.9$ and reproduces the basic properties of the EOS with the NL3 parameter set at the same time. )

\begin{table}[ht]
\begin{center}
\begin{tabular}{lccccc} \hline \hline
            & B & D & K & BD &  \\ \hline
${g_{\sigma}}^2/{m_\sigma}^2$  & 395.474 & 384.641 & 388.669 & 407.537 & (GeV$^{-2}$) \\  \hline 
${g_{\omega}}^2/{m_\omega}^2$  & 259.581 & 250.491 & 252.824 & 278.194 &(GeV$^{-2}$) \\  \hline 
${c_{\sigma 3}}/{g_\sigma}^3$  & 2.2644 & 2.3837 & 2.4469 & 1.4938 &(MeV) \\  \hline 
${c_{\sigma 4}}/{g_\sigma}^4$  & $-0.37906 $ & $-0.46132$ & $-0.54749$ & $-0.12085$ & $\times 10^{-2}$  \\  \hline 
${c_{\omega 4}}/{g_\omega}^4$  & $-0.14$ & 0.0 & 0.0 & $-0.37744$ & $\times 10^{-2}$ \\  \hline
${g_{\omega 3}}/{g_\omega}^3$  & 0.0 & 0.2 & 0.0 & $-0.35$ & (GeV$^{-2}$) \\  \hline 
${c_{\sigma 2\omega 2}}/({g_\sigma}{g_\omega})^2$  & 0.0 & 0.0 & $0.0015$ & 0.0 &  \\  \hline
\hline
\end{tabular}
\end{center}
\caption{The parameter sets B,D,K and BD. All parameter sets reproduce the basic properties of the EOS with the parameter set NL3.}
\label{Table 1}
\end{table}


In Figs. 3$\sim$5, we show ${m_\omega^*}^2$, $g_{\rm v\omega}^*$ and $\epsilon /\rho_{\rm B}$ as functions of the baryon density. 
In each case, the EOS becomes stiffer than the EOS with the original NL3 parameter set because of the effective $\omega$-meson mass reduction. 
In the cases of the set D, the enhancement of the effective $\omega$-nucleon couplings also contributes the stiffness of the EOS. 

If we require that the $\omega$-meson mean field is proportional to the baryon density, there should be the following relation between $g_{\omega 3}$ and $c_{\omega 4}$. 
\begin{eqnarray}
c_{\omega 4}=3{g_{\omega 3}m_\omega^2\over{g_\omega}}
\label{eq:E200}
\end{eqnarray}
We call this parameter set a "set BD". 
In this parameter set, we determine $g_{\omega 3}$ and $c_{\omega 4}$ to satisfy ${m_\omega^*}^2/m_\omega^2=0.9$ and the condition above. 
The numerical results are shown in Figs. 3$\sim$5. 
In spite of the effective $\omega$-meson mass reduction, the EOS becomes softer because of the reduction of the effective $\omega$-nucleon coupling. 


Next we examine the effects of the explicit density dependence in the effective interactions.  
In parameter set ED1, $g_{\rm v\omega}^*$ decreases as density increases because of the effective interaction $\bar{\psi}g_{\omega V2}\omega V^2\gamma_0\psi$. 
On the other hand, the effective $\sigma$-nucleon interaction also decreases as density increases because of the effective interaction $\bar{\psi}g_{\sigma S}\sigma S\psi$. 
In parameter set ED2, an effective interaction $\bar{\psi}g_{\omega 2V}\omega^2 V\gamma_0\psi$ is added as well as $\bar{\psi}g_{\omega V2}\omega V^2\gamma_0\psi$ and $\bar{\psi}g_{\sigma S}\sigma S\psi$. 
This new interaction makes the effective $\omega$-meson mass smaller if $g_{\omega 2V}$ is positive. 
On the other hand, this term makes the effective $\omega$-nucleon coupling larger. 

\begin{table}[ht]
\begin{center}
\begin{tabular}{lcccccc} \hline \hline
            & ED1 & ED2 & QSR & \\ \hline
${g_{\sigma}}^2/{m_\sigma}^2$  & 410.299 & 412.523 & 444.050 & \\  \hline 
${g_{\omega}}^2/{m_\omega}^2$  & 290.124 & 291.622 & 327.489 & \\  \hline 
${c_{\sigma 3}}/{g_\sigma}^3$  & 0.0 & 0.0 & 0.95677 & (MeV) \\  \hline 
${c_{\sigma 4}}/{g_\sigma}^4$  & 0.0 & 0.0 & 0.16515 & $\times 10^{-2}$ \\ \hline 
${g_{\sigma S}}/{g_\sigma}$  & $-92.350$ & -95.441 & 0.0 & (GeV$^{-3}$) \\ \hline 
${g_{\omega V2}}/{g_\omega}$  & $-4.4088$ & -16.617 & 0.0 & 
$\times 10^4$(GeV$^{-6}$) \\  \hline 
${g_{\omega 2 V}}/{g_\omega}^2$  & 0.0 & 270.0 & 0.0 & (GeV$^{-4}$) \\  \hline 
$C/{g_\omega}^2$  & 0.0 & 0.0 & 0.48576 & (GeV$^{-1}$) \\  \hline
\hline
\end{tabular}
\end{center}
\caption{The parameter sets ED1, ED2 and QSR. All parameter sets reproduce the basic properties of the EOS with the parameter set NL3.}
\label{Table 2}
\end{table}

In Figs. 6$\sim$8, we show ${m_\omega^*}^2$, $g_{\rm v\omega}^*$ and $\epsilon/\rho_{\rm B}-m$ as functions of the baryon density for the parameter sets ED1 and ED2. 
In the parameter set ED2, the $g_{\omega 2V}$ is determined to satisfy ${m_\omega^*}^2/m_\omega^2=$0.9. 
Because of the reduction of the effective $\omega$-meson mass and the enhancement of the effective $\omega$-nucleon coupling, the EOS with the set ED2 becomes stiffer than the one with the set ED1. 

We remark that the case with the parameter set ED2 is different from the case with the parameter set BD which we have discussed before. 
In the case of the parameter set BD, the $\omega$-meson self-interaction term affects the effective $\omega$-meson mass more strongly than the multi $\omega$-nucleon interaction, because the effective $\omega$-meson mass is defined as a second derivative of the energy density with respect of the $\omega$-meson field and the term which is proportional to $\omega^4$ is more important for the effective $\omega$-meson mass than the term which is proportional to $\omega^3$. 
Therefore, we can make the effective $\omega$-meson mass smaller and the effective $\omega$-nucleon couplings smaller at the same time using the negative coefficient $c_{\omega 4}$. 
To the contrary, in the case with the parameter set ED2, the interaction term 
$\bar{\psi}g_{\omega 2V}\omega^2V\gamma_0\psi$ affects the effective $\omega$-meson mass more strongly than the corresponding mesonic potential term 
$-{1\over{2}}g_{\omega 2V}\omega^2V^2$. 
Therefore, we should make $g_{\omega 2V}$ positive to decrease the effective $\omega$ meson mass and this makes the effective $\omega$-nucleon coupling larger. 
In general, the term which includes higher interactions of $\omega$-meson is more important for the effective $\omega$-meson mass. 
In this viewpoint, the difference between the $\omega$-meson field and the auxiliary field $V$ is crucial and important.


The QCD sum-rule has predicted that the $\omega$-meson mass decreases when the density increases. \cite{rf:Hatsuda} 
At low density, it is approximated by 
\begin{eqnarray}
m^*_\omega (\rho_{\rm B} )=m_\omega -{0.18m_\omega\over{\rho_{\rm B0}}}\rho_{\rm B}. 
\label{eq:E83}
\end{eqnarray}
This gives 
\begin{eqnarray}
\Delta {{m}^*_\omega}^2\equiv \tilde{m}_\omega^2-m_\omega^2
=\left(m_\omega -{0.18m_\omega\over{\rho_{\rm B0}}}\rho_{\rm B}\right)^2-m_\omega^2\sim -2C\rho_{\rm B}, 
\label{eq:E84}
\end{eqnarray}
where $C=0.18m_\omega^2/\rho_{\rm B0}$. 
This effective mass is defined as a pole of the propagator and is different from our effective mass which is defined as a meson self-energy at the zero external momentum. 
However, in this paper, we neglect this difference, use Eq. (\ref{eq:E84}) as a first approximation and try to embed it in the hadron field theory. 

There are two difficulties in this trial. 
If we use the solution (\ref{eq:E12a})$\sim$(\ref{eq:E15}), we could not embed Eq. (\ref{eq:E84}) directly into the mesonic potential because of the condition $u_{0,1}=0$. 
Therefore, we use the higher-order interaction of $\omega$-meson to include the linear density dependence in Eq. (\ref{eq:E84}). 
Furthermore, the property for the boost transformation is already unclear in (\ref{eq:E83}) and it is difficult to determine the appropriate form of the multi $\omega$-nucleon couplings. 
Here, we naively add the terms to the Lagrangian. 
\begin{eqnarray}
\Delta \Sigma_{\rm v}=C\omega^2~~~~~{\rm and}~~~~~\Delta U_{\rm MV}={2Cm_\omega^2\over{3g_\omega}}\omega^3\label{eq:E85}
\end{eqnarray}
Both of these two terms are needed because we require that the $\omega$-meson 
 mean field should be proportional to the baryon density. 
This requirement may be natural in the viewpoint of the quark physics. \cite{rf:Kunihiro} 
These terms gives negative contribution $-2C\rho_{\rm B}$ to ${m_\omega^*}^2$. 
On the other hand, it also makes the effective $\omega$-nucleon couplings smaller. 
We call this parameter set QSR. 


In Figs. 6$\sim$8, we show ${m_\omega^*}^2$, $g_{\rm v\omega}^*$ and $\epsilon /\rho_{\rm B}-m$ as functions of the baryon density for the parameter set QSR. 
As is in the case of the parameter set BD, in spite of the $\omega$-meson mass reduction, the EOS becomes softer because of the reduction of the effective $\omega$-nucleon coupling. 


\section{Effective Nucleon-Nucleon Interactions}

If we solve the equations of motion for meson mean-field, $\sigma$ and $\omega$ are written in terms of $\rho_{\rm s}$ and $\rho_{\rm B}$. 
In fact, in the original Walecka model, $\sigma$ and $\omega$
are removed by 
\begin{eqnarray}
\sigma ={g_\sigma\over{m_\sigma^2}}\rho_{\rm s}
~~~~~{\rm and}~~~~~
\omega ={g_\omega\over{m_\omega^2}}\rho_{\rm B}.
\label{eq:E89}
\end{eqnarray}
Therefore, if $S(=\rho_{\rm s})$ is written by $V(=\rho_{\rm B})$, 
$\Sigma_{\rm s}$, $\Sigma_{\rm v}$ and $U_{\rm M}$ are functions which depend 
$V$ only. 

Barreiro \cite{rf:Barreiro} has tried to remove $\sigma$ and $\omega$ using the relation 
\begin{eqnarray}
\sigma ={\Sigma_{\rm s}\over{g_\sigma}}
~~~~~{\rm and}~~~~~
\omega ={\Sigma_{\rm v}\over{g_\omega}}
\label{eq:E90}
\end{eqnarray}
rather than Eq. (\ref{eq:E89}). 
He showed that, combining the QCD sum-rule results \cite{rf:Cohen} 
\begin{eqnarray}
\Sigma_{\rm s}=-{\sigma_Bm\over{m_\pi^2f_\pi^2}}\rho_B
~~~~~{\rm and}~~~~~
\Sigma_{\rm v}=-{8m_{\rm q}m\over{m_\pi^2f_\pi^2}}\rho_B
\label{eq:E91}
\end{eqnarray}
with Bonn potential, \cite{rf:Machleidt} the saturation of nuclear matter is well described. 
However, his method does break the thermodynamical consistency. 
Instead of using (\ref{eq:E90}), we use the evolution equation (\ref{eq:E10}) with Eqs. (\ref{eq:E91}). 
Since $\Sigma_{\rm s}$, 
$\Sigma_{\rm v}$ and $U_{\rm M}$ depend only $V$, the evolution equation  ({\ref{eq:E10}) reduces to 
\begin{eqnarray}
{\partial \epsilon\over{\partial V}}=0. 
\label{eq:E92}
\end{eqnarray}


Using this equation, $U_{\rm M}$ is determined as
\begin{eqnarray}
U_{\rm M}=-{1\over{2}}{8m_{\rm q}m\over{m_\pi^2f_\pi^2}}\rho_{\rm B}
+{\sigma_{\rm B}m\over{m_\pi^2f_\pi^2}}\int_0^{\rho_{\rm B}} \rho_{\rm s}(\rho_{\rm B}^\prime )d\rho_{\rm B}^\prime,  
\label{eq:E93}
\end{eqnarray}
where $m_\pi$, $f_\pi$, $m_q$ and $\sigma_{\rm B}$ are 
$\pi$-meson mass, the pion decay constant, the current quark mass and the nucleon sigma-term, respectively. 
Using $\rho_{\rm s}\sim \rho_{\rm B}$, we obtain
\begin{eqnarray}
U_{\rm M}&=& -{1\over{2}}{8m_{\rm q}m\over{m_\pi^2f_\pi^2}}\rho_{\rm B}^2
+\int_0^{\rho_{\rm B}}{\sigma_Bm\over{m_\pi^2f_\pi^2}}\rho_{\rm s}^\prime
{d\rho_{\rm B}^\prime \over{d\rho_{\rm s}^\prime}}d\rho_{\rm s}^\prime 
\nonumber\\
&\sim & 
-{1\over{2}}{8m_{\rm q}m\over{m_\pi^2f_\pi^2}}\rho_{\rm B}^2
+\int_0^{\rho_{\rm B}}{\sigma_Bm\over{m_\pi^2f_\pi^2}}\rho_{\rm s}^\prime
d\rho_{\rm s}^\prime 
\nonumber\\
&=& -{1\over{2}}{8m_{\rm q}m\over{m_\pi^2f_\pi^2}}\rho_{\rm B}^2
+{1\over{2}}{\sigma_Bm\over{m_\pi^2f_\pi^2}}\rho_{\rm s}^2
\label{eq:E94}
\end{eqnarray}
Comparing this equations with 
the potential 
\begin{eqnarray}
U_{\rm M}
=-{1\over{2}}{g_\omega^2\over{m_\omega^2}}\rho_{\rm B}^2
+{1\over{2}}{g_\sigma^2\over{m_\sigma^2}}\rho_{\rm s}^2
\label{eq:E95}
\end{eqnarray}
in the original Walecka model, 
we obtain 
\begin{eqnarray}
{g_\omega^2\over{m_\omega^2}}={8m_{\rm q}m\over{m_\pi^2f_\pi^2}}
~~~~~{\rm and}~~~~~
{g_\sigma^2\over{m_\sigma^2}}\sim {\sigma_Bm\over{m_\pi^2f_\pi^2}}. 
\label{eq:E96}
\end{eqnarray}
Since 
${g_\sigma^2\over{m_\sigma^2}}=303$ and ${g_\omega^2\over{m_\omega^2}}=222$
in the original Walecka model, Eq. (\ref{eq:E96}) is satisfied for 
$m_\pi=$138MeV, $f_\pi=$93MeV, $m_q=$4.8MeV and $\sigma_{\rm B}=$53MeV. 
Because $\rho_{\rm s}=<\bar{\psi}\psi >$ and $\rho_{\rm B}=<\bar{\psi}\gamma_0\psi >$, the potential $U_{\rm M}$ may be regarded as the effective nucleon-nucleon 
interaction caused by the meson exchange between nucleons. 
In fact, in the chiral perturbation theory, \cite{rf:Vretenar} Eq. (\ref{eq:E94}) is naturally derived from the effective nucleon-nucleon interactions. 

In Fig. 9, using the second equation of Eq. (\ref{eq:E91}) with $m_\pi =138$MeV, $f_\pi=93$MeV, $m_{\rm q}=7$MeV and $\sigma_{\rm B}=45$MeV, \cite{rf:Gasser} 
we show $\Sigma_{\rm v}$ as a function of baryon density. 
In Fig. 10, using the exact potential (\ref{eq:E93}) with the parameters above, we show $\epsilon/\rho_{\rm B}-m$ as a function of baryon density. 
The $\epsilon/\rho_{\rm B}-m$ is large and does not satisfy the saturation conditions. 


Next, we try to embed the $\omega$-meson mass reduction, which is predicted by the QCD sum rule, \cite{rf:Hatsuda} into $\Sigma_{\rm v}$. 
As is in the case of the parameter set QSR, we add the term
\begin{eqnarray}
\Delta \Sigma_{\rm v}=C\omega^2~~~~~{\rm and}~~~~~\Delta U_{\rm MV}={2Cm_\omega^2\over{3g_\omega}}\omega^3\label{eq:E97}
\end{eqnarray}
to $\Sigma_{\rm v}=-g_\omega\omega$ and the vector part of the potential $U_{\rm M}$ in the original Walecka model. 
In this modification, the equation of motion for the $\omega$-meson is not changed and reads 
\begin{eqnarray}
\omega ={g_\omega\over{m_\omega^2}}\rho_{\rm B}. 
\label{eq:E98}
\end{eqnarray}
Putting (\ref{eq:E98}) into the total $\Sigma_{\rm v}$, we obtain
\begin{eqnarray}
\Sigma_{\rm v}&=&-g_\omega\omega+C\omega^2
= -{g_\omega^2\over{m_\omega^2}}\rho_{\rm B}\left(1-{C\over{m_\omega^2}}\rho_{\rm B}\right)
\nonumber\\
&=&-{m_{\rm q}m\over{m_\pi^2f_\pi^2}}\rho_{\rm B}\left(1-0.18{\rho_{\rm B}\over{\rho_{\rm B0}}}\right),  
\label{eq:E99}
\end{eqnarray}
where we have used the first relation in Eq. (\ref{eq:E96}). 
The vector part of the potential becomes 
\begin{eqnarray}
U_{\rm MV}&=&-{1\over{2}}{g_\omega^2\over{m_\omega^2}}\rho_{\rm B}^2
+{2g_\omega^2\over{3m_\omega^4}}C\rho_{\rm B}^3
=-{1\over{2}}{g_\omega^2\over{m_\omega^2}}\rho_{\rm B}^2
\left(1-0.24{\rho_{\rm B}\over{\rho_{\rm B0}}}\right)
\nonumber\\
&=&-{1\over{2}}{8m_{\rm q}m\over{m_\pi^2f_\pi^2}}\rho_{\rm B}^2
\left(1-0.24{\rho_{\rm B}\over{\rho_{\rm B0}}}\right). 
\label{eq:E100}
\end{eqnarray}
In Fig. 9 and 10, using Eqs. (\ref{eq:E99}) and (\ref{eq:E100}), we show $\Sigma_{\rm v}$ and $\epsilon/\rho_{\rm B}-m$ as a function of baryon density. 
In spite of the enhancement of $U_{\rm MV}$, the equation of state becomes much softer because the absolute value of $\Sigma_{\rm v}$ becomes much smaller. 
We remark that these results depend strongly on the values of $m_{\rm q}$ and $\sigma_{\rm B}$. 
For example, if we use $m_{\rm q}=5.5$MeV instead of $m_{\rm q}=7$MeV, 
$\epsilon/\rho_{\rm B}-m$ becomes negative at high densities. 
In conclusion, the total effect of the reduction of the $\omega$-meson mass is attractive. 


\section{Summary}

In summary, we have studied the relation among the effective hadron mass, the 
effective interactions and the EOS by using the generalized relativistic mean field theory. 
The results obtained in this paper is summarized as follows. 

(1) The reduction of the effective $\omega$-meson mass stiffens the EOS. 
 
(2) In many cases, the reduction of the $\omega$-meson mass is followed by the 
enhancement of the $\omega$-nucleon interaction which makes EOS more stiff. 

(3) However, if we require that the $\omega$-meson mean field is proportional to the baryon density, the reduction of the $\omega$-meson mass induces the reduction of the effective $\omega$-nucleon couplings because of the simple relation between the effective $\omega$-meson mass and the effective $\omega$-nucleon coupling. 
Consequently, the EOS becomes softer. 
The total effect of the reduction of the $\omega$-meson mass is attractive. 

In our analyses, the difference between the meson field $\omega$ and the auxiliary field $V$ is crucial and important. 
In the nucleon, there may be a quasi field which consists of the quark and anti-quark and couples with a quark itself. 
If the quasi field comes out of the nucleon, they becomes a "true" meson by the confinement of the quark and anti-quark. 
The confinement is a very complicated and drastic phenomenon. 
However, there may be continuity in some degree between the quasi field and the meson, because of the conservation of the baryon number. 
Since the quark and the anti-quark are confined, the mesons become "physical" particles and it is meaningful to consider the masses of them. 
On the other hand, $V$ is nothing but an auxiliary field which is introduced into the theory for convenience. 
It is regarded as the one which represents the effect of the effective nucleon-nucleon interactions. 
Since the nucleons are not confined, $V$ could not be a "physical" particle and there is no physical meaning for its masses. 
In this viewpoint, we should distinguish the $\omega$-meson from the auxiliary field $V$. 
Furthermore, the result of the quark physics may be more naturally described by the higher order $\omega$-meson interactions rather than the explicit density dependence, since the $\omega$-meson field consists of the quark and the anti-quark, and is more directly connected with the quark physics inside the nucleons than the auxiliary field $V$. 

We remark that the effective $\omega$-meson mass which we discussed in this paper is somewhat different from the one defined by using the pole of the full propagator of the meson. 
As is discussed by Kurasawa and Suzuki, \cite{rf:Kurasawa} the vacuum contribution is important for such a "pole mass". 
If we study the "pole mass", we must include the effective interactions which include derivative couplings in Eq. (\ref{eq:E1}). 
They are open problems in future. 

Our model with the P.S. BD may be too simple. 
However, we emphasize that the equation (\ref{eq:E67c}) is very general and is applicable to more realistic model. 
One of the most interesting model for the nuclear matter with the chiral symmetry is the extended chiral sigma model. \cite{rf:Boguta2,rf:Ogawa} 
This model seems very reasonable and very interesting, but it is indicated that the EOS of this model becomes very hard. \cite{rf:Ogawa} 
In this model, the $\sigma$-$\omega$ mixing interactions appears in the resulting $\omega$-meson potential as follows. 
\begin{eqnarray}
U_\omega = {1\over{2}} m_\omega^2\omega_\mu\omega^\mu -m_\omega^2 {\sigma\over{f_\pi}}\omega_\mu\omega^\mu
+ {1\over{2}} m_\omega^2 {\sigma^2\over{f_\pi^2}}\omega_\mu\omega^\mu. 
\label{eq:E0826a}
\end{eqnarray}
The effective $\omega$-meson mass is given by \cite{rf:Ogawa}
\begin{eqnarray}
m_\omega^*=m_\omega\left( 1-{\sigma\over{f_\pi}}\right). 
\label{eq:E0826b}
\end{eqnarray}
Therefore the effective $\omega$-meson mass becomes smaller as the value the $\sigma$-meson mean field becomes larger and the EOS becomes stiff. 
In this model, the $\omega$-meson mean field is not proportional to the baryon density because the $\sigma$-$\omega$ mixing interactions in (\ref{eq:E0826a}). 
If we required the $\omega$-meson mean field is proportional to the baryon density, the interactions 
\begin{eqnarray}
\bar{\psi}\{g_\omega(-{2\over{f_\pi}}\sigma+{1\over{f_\pi^2}}\sigma^2)\omega_\mu \gamma^\mu \}\psi. 
\label{eq:E0826c}
\end{eqnarray}
should be added to the the ordinary $\omega$-nucleon coupling $\bar{\psi}g_\omega\omega_\mu\gamma^\mu\psi$. 
In this case, the condition (\ref{eq:E67c}) is satisfied. 
Therefore, it is expected that the EOS becomes softer, since the effective $\omega$-nucleon coupling becomes smaller as the effective $\omega$-meson mass becomes smaller. 
We also remark that the mixing mass term ${\partial^2 \epsilon\over{\partial\sigma\partial\omega}}$ vanishes due to the terms in Eq. (\ref{eq:E0826c}). 
The $\sigma$-$\omega$ mixing interactions (\ref{eq:E0826c}) are obtained if there is a term 
\begin{eqnarray}
-\bar{\Psi}\{{g_\omega\over{f_\pi^2}} ({\sigma^\prime}^2+\pi_a^2)\omega_\mu \gamma^\mu \}\Psi
\label{eq:E0827a}
\end{eqnarray}
in the initial chiral sigma model Lagrangian, 
where $\Psi$ and $\sigma^\prime$ are the original nucleon and $\sigma$-meson fields, respectively, and $\pi_a$ is the pion field. 

In this paper, the generalized relativistic mean field theory is improved by including the baryon density and the scalar density as auxiliary bosonic fields. 
The method may be justified and improved by the auxiliary field method in the field theory. \cite{rf:Gross,rf:Kugo,rf:Kikkawa,rf:Kashiwa} 
The baryon density or the baryonic chemical potential flow equation may be useful in the nuclear physics. 

\bigskip

\centerline{\bf Acknowledgement}

\bigskip

Authors thank to A. Hasegawa and M. Nakano for useful discussions. 
One (H.K.) of the authors also thanks to T. Kunihiro for useful discussions and suggestions, and his encouragement. 
Without his crucial suggestion we could not complete this work. 

\bigskip


\vfill\eject

\center{\bf Figure Captions} 

\begin{flushleft}

\bigskip

Fig. 1  The diagrammatic descriptions for the nucleon self-energies (a), 
the effective meson-nucleon couplings (b), the equation of motion for meson (c) and the meson self-energies (d). 

\bigskip

Fig. 2 At the normal density, $K$ is shown as a function of the effective nucleon mass $m_0^*$. 
The various curves represents results with the fixed values of 
$-^t\hat{\bf g}({M^*}^2)^{-1}\hat{\bf g}$ ( shown in GeV$^{-2}$ ) and with $\rho_{\rm B0}=0.148$fm$^{-3}$. 
The results for the well-known parameter sets NL1 \cite{rf:Reinhard}, NL-SH \cite{rf:Sharma}, TM1 \cite{rf:Sugahara} and NL3 \cite{rf:Lalazissis} are also shown.

\bigskip

Fig. 3 The ${m_\omega^*}^2/m_\omega^2$ is shown as a function of baryon density. 
In the case with the parameter set NL3, ${m_\omega^*}^2/m_\omega^2=1$. 

\bigskip

Fig. 4 The $g_{\rm v\omega}^*/g_\omega$ is shown as a function of baryon density. 
In the cases with the parameter sets NL3, B and K, $g_{\rm v\omega}^*/g_\omega =1$. 

\bigskip

Fig. 5 The $\epsilon /\rho_{\rm B}-m$ (in MeV) is shown as a function of baryon density. 
\bigskip

Fig. 6 The ${m_\omega^*}^2/m_\omega^2$ is shown as a function of baryon density. 
In the case with the parameter sets NL3 and ED1, ${m_\omega^*}^2/m_\omega^2=1$. 
\bigskip

Fig. 7 The $g_{\rm v\omega}^*/g_\omega$ is shown as a function of baryon density. 
In the cases with the parameter set NL3, $g_{\rm v\omega}^*/g_\omega =1$. 

\bigskip

Fig. 8 The $\epsilon /\rho_{\rm B}-m$ (in MeV) is shown as a function of baryon density. 
\bigskip

Fig. 9 The $\Sigma_{\rm v}$ (in GeV) is shown as a function of baryon density. 
The solid, dashed and dash-dotted curves represent the results with $m_{\rm q}=7$MeV and no $\omega$-meson mass modification, the result with $m_{\rm q}=7$MeV and the $\omega$-meson mass modification and the result with $m_{\rm q}=5.5$MeV and the $\omega$-meson mass modification, respectively. 

\bigskip

Fig. 10 The $\epsilon /\rho_{\rm B}-m$ (in MeV) is shown as a function of baryon density. 
The various curves have the same notation as in Fig. 9. 

\end{flushleft}


\vfill\eject
\oddsidemargin 0mm
\evensidemargin 0mm
\textwidth 190mm
\topmargin 0mm
\textheight 250mm
\headsep -5mm
\topskip -5mm

\begin{center}
\begin{center}
\begin{tabular}{c}
\includegraphics[height=200mm,width=160mm] {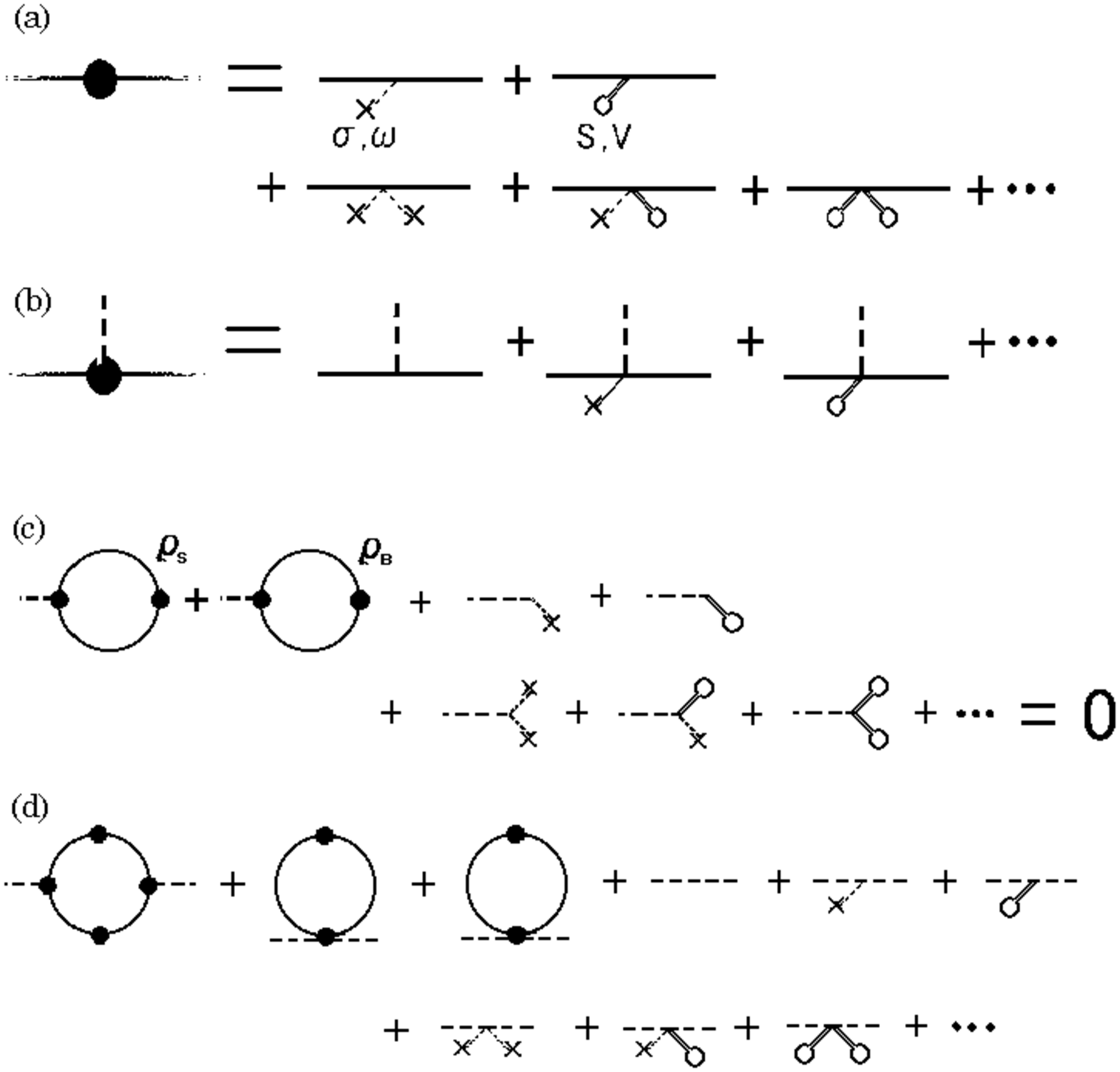} \\ 
Fig.1
\end{tabular}
\end{center}
\newpage
\begin{center}
\begin{tabular}{c}
\includegraphics[height=80mm,width=80mm] {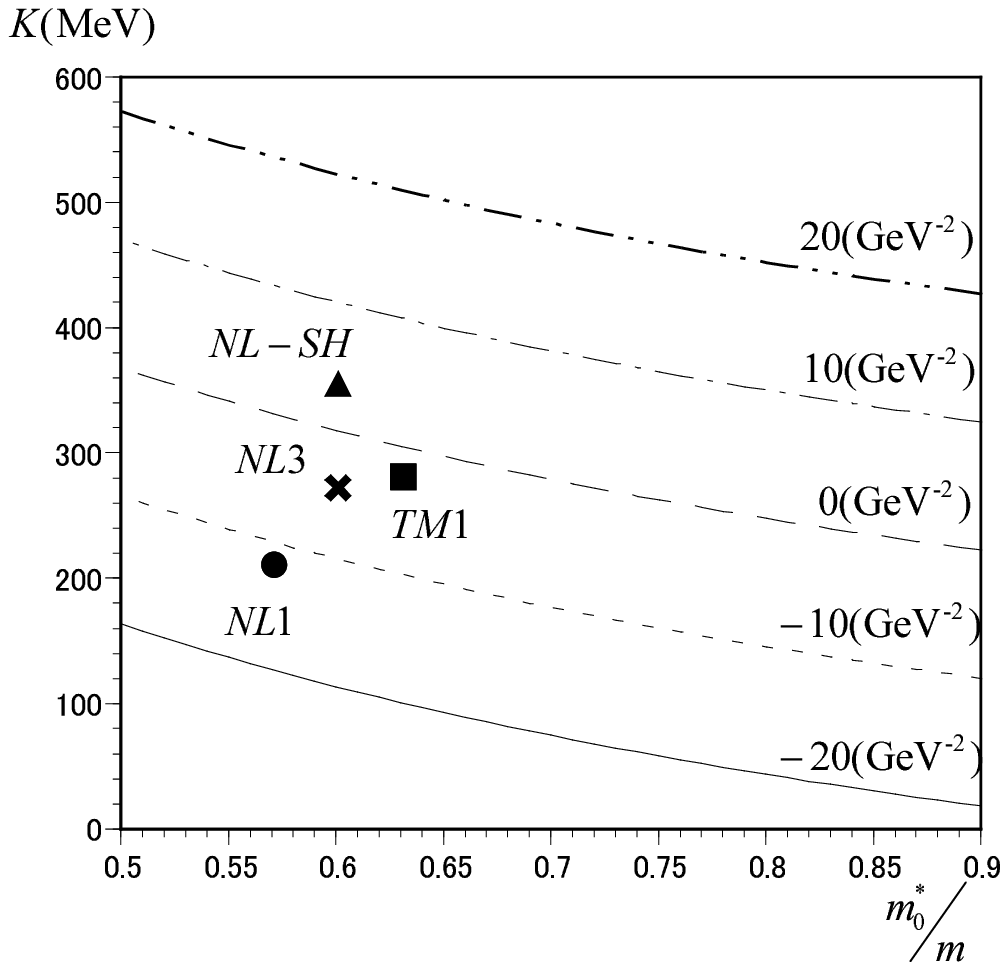} \\
Fig.2
\end{tabular}
\end{center}

\begin{center}
\begin{tabular}{cc}
\includegraphics[height=70mm,width=80mm] {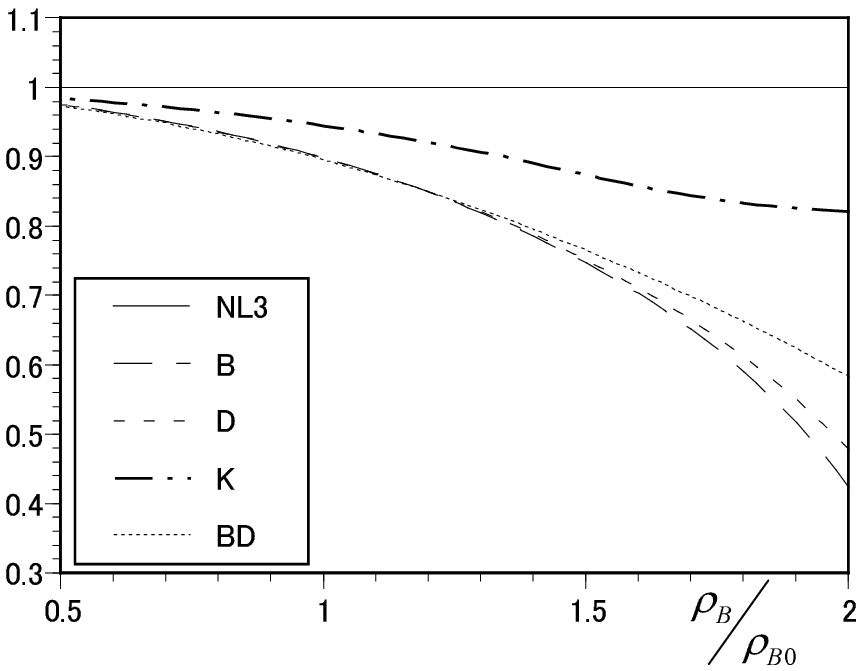} &
\includegraphics[height=70mm,width=80mm] {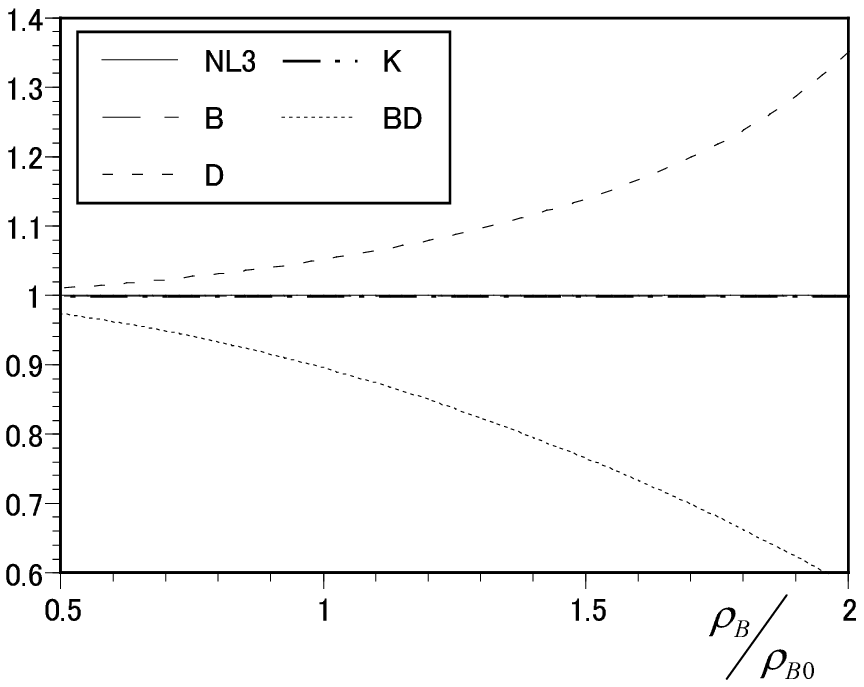}\\

Fig.3
&
Fig.4
\end{tabular}
\end{center}
\begin{center}
\begin{tabular}{cc}
\includegraphics[height=70mm,width=80mm] {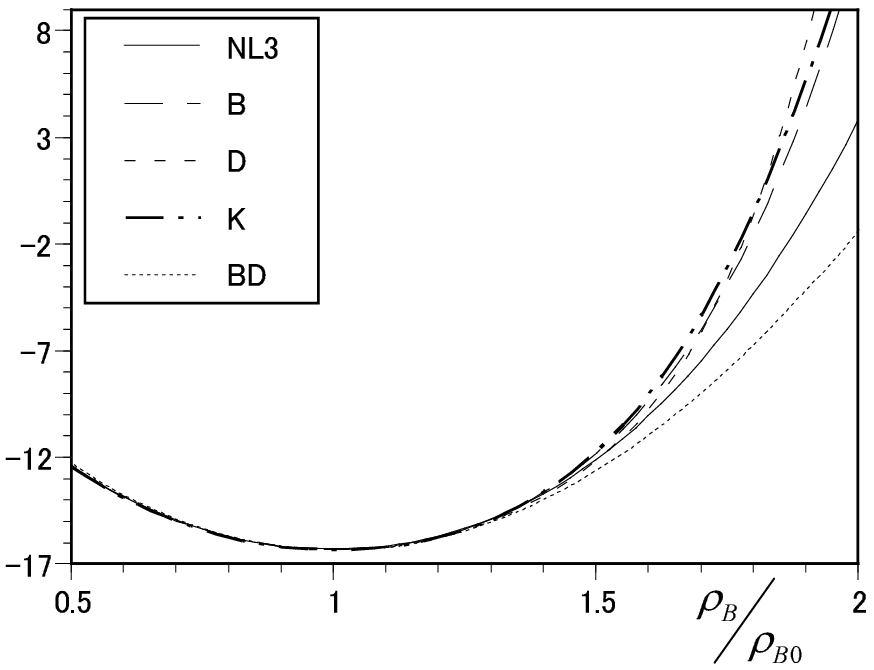} \\ 

Fig.5
\end{tabular}
\end{center}

\newpage
\begin{center}
\begin{tabular}{cc}
\includegraphics[height=70mm,width=80mm] {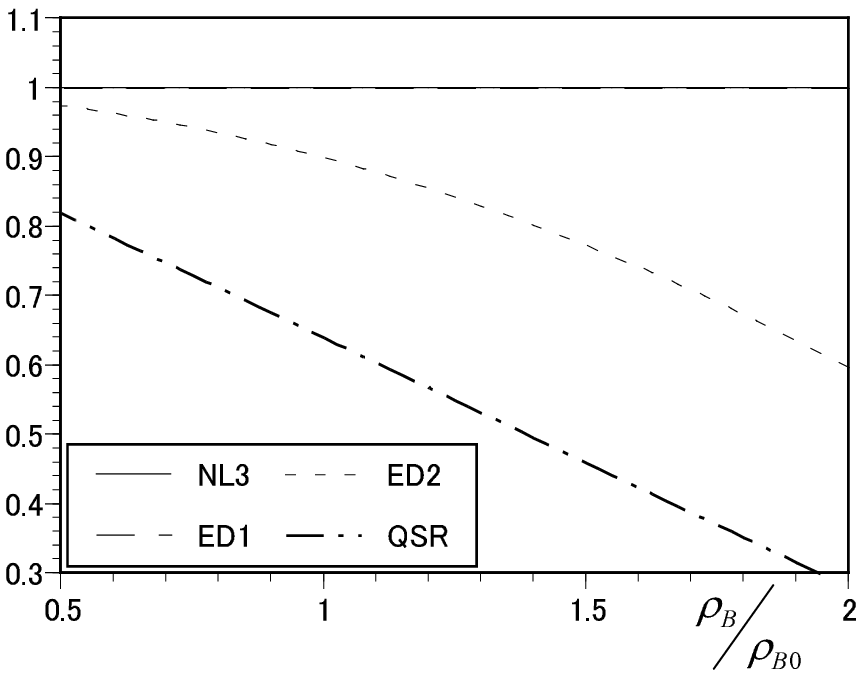} &
\includegraphics[height=70mm,width=80mm] {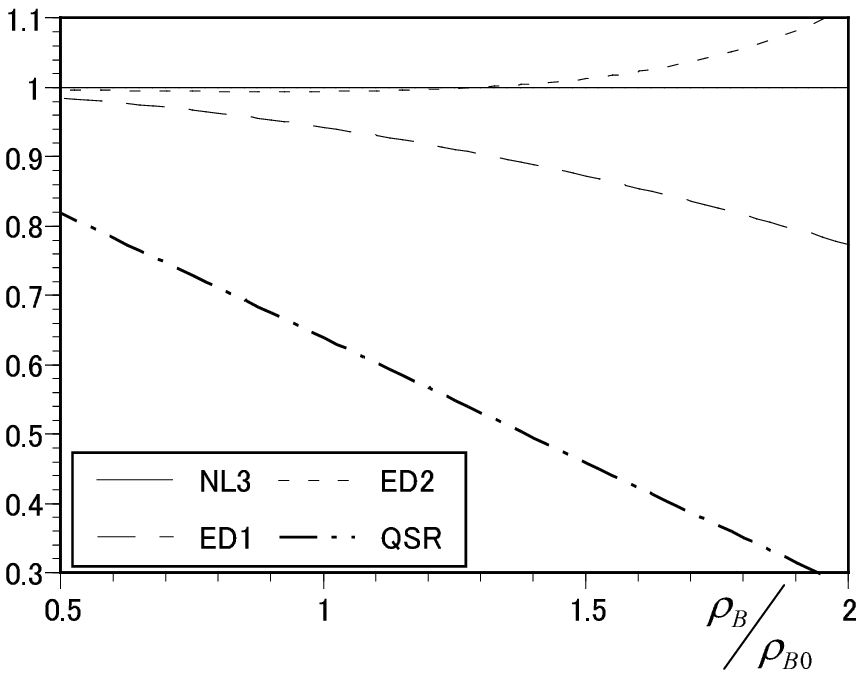}\\

Fig.6
&
Fig.7
\end{tabular}
\end{center}
\begin{center}
\begin{tabular}{cc}
\includegraphics[height=70mm,width=80mm] {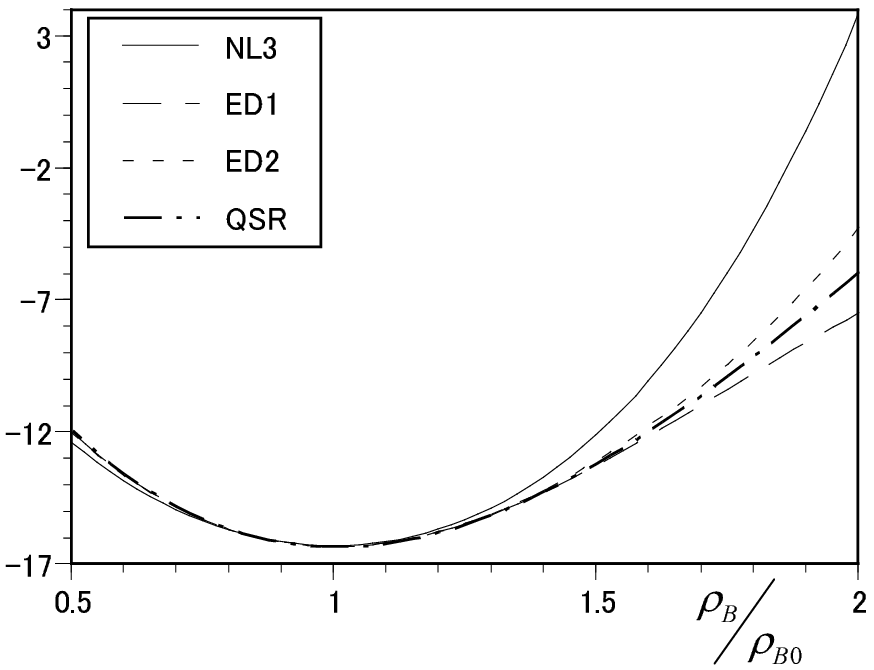} \\ 

Fig.8
\end{tabular}
\end{center}

\begin{center}
\begin{tabular}{cc}
\includegraphics[height=70mm,width=80mm] {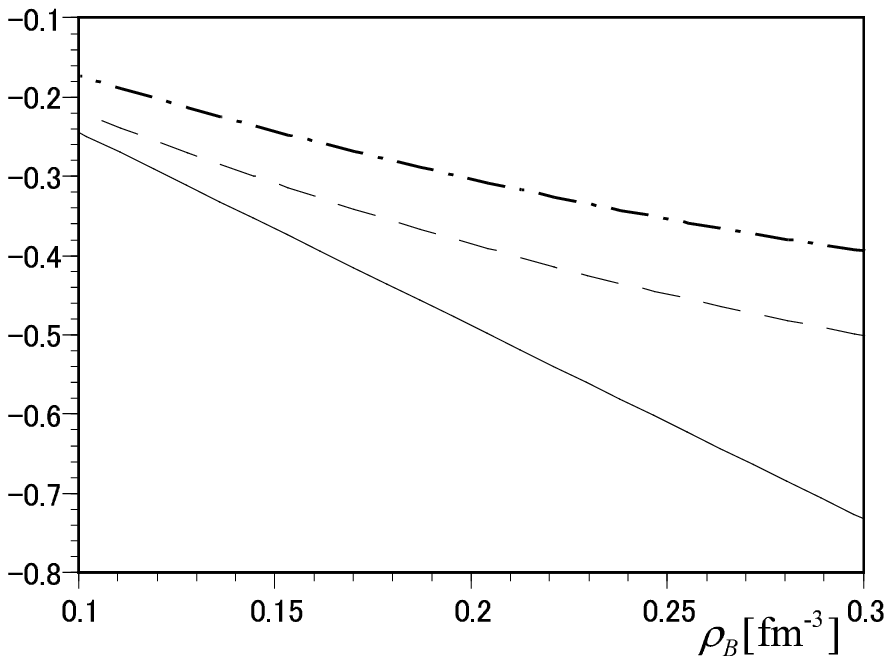} &
\includegraphics[height=70mm,width=80mm] {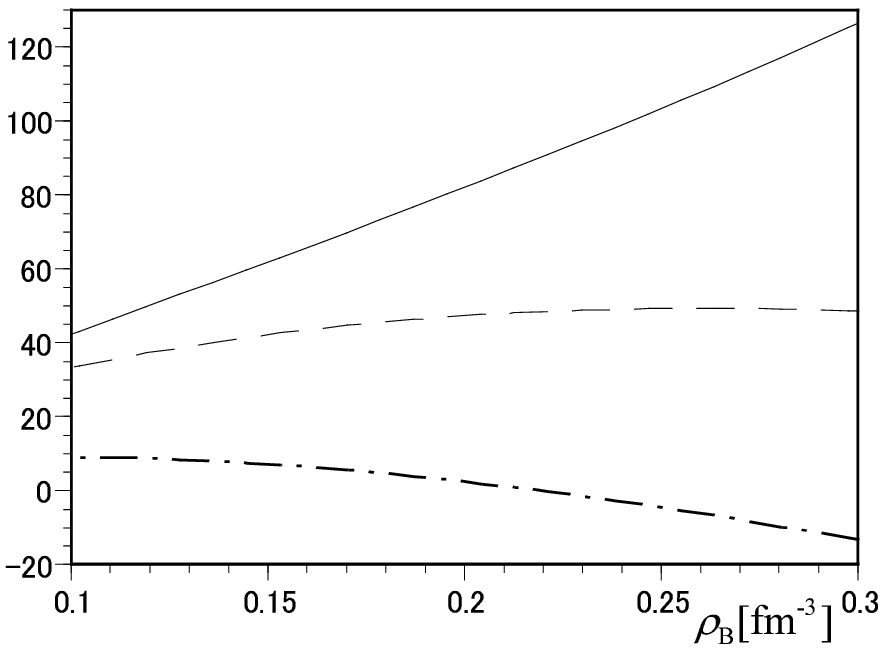}\\

Fig.9
&
Fig.10
\end{tabular}
\end{center}

\end{center}
\end{document}